\documentclass[%
 reprint,onecolumn,notitlepage,nofootinbib,
 amsmath,amssymb,
 aps,
]{revtex4-1}

\usepackage{graphicx}
\usepackage{subcaption}
\usepackage[justification=justified, singlelinecheck=off, compatibility=false]{caption}
\usepackage{amsmath}
\usepackage{bm}
\usepackage{amssymb}
\usepackage{xcolor}
\usepackage{times}
\usepackage{float}
\usepackage{hyperref}
\usepackage[utf8]{inputenc}
\hypersetup{
    colorlinks=true,       
    linkcolor=red,          
    citecolor=blue,        
    filecolor=magenta,      
    urlcolor=blue           
}

\newcommand{\eref}[1]{Eq.~(\ref{#1})}
\newcommand{\fref}[1]{Fig.~\ref{#1}}
\newcommand{\sref}[1]{Sec.~\ref{#1}}


\begin{document}

\title{Using Secondary Tau Neutrinos to Probe Heavy Dark Matter Decays in Earth}

\date{\today}

\author{Matthew Saveliev}
\affiliation{Department of Physics \& Astronomy, Bowdoin College, Brunswick, Maine 04011}

\author{Jeffrey Hyde}
\email{jhyde@bowdoin.edu}
\affiliation{Department of Physics \& Astronomy, Bowdoin College, Brunswick, Maine 04011}

\begin{abstract}
Dark matter particles can be gravitationally trapped by celestial bodies, motivating searches for localized annihilation or decay. If neutrinos are among the decay products, then IceCube and other neutrino observatories could detect them. We investigate this scenario for dark matter particles above $m_{\chi} \gtrsim$ PeV producing tau neutrino signals, using updated modeling of dark matter capture and thermalization. At these energies, tau neutrino regeneration is an important effect during propagation through Earth, allowing detection at distances far longer than one interaction length. We show how large energy loss of tau leptons above $\sim$ PeV drives a wide range of initial energies to the same final energy spectrum of ``secondary" tau neutrinos at the detector, and we provide an analytic approximation to the numerical results. This effect enables an experiment to constrain decays that occur at very high energies, and we examine the reach of the IceCube high-energy starting event (HESE) sample in the parameter space of trapped dark matter annihilations and decays above PeV. We find that the parameter space probed by IceCube searches would require dark matter cross sections in tension with existing direct-detection bounds.
\vspace{1cm}
\end{abstract}

\maketitle

\setlength{\parindent}{20pt}

\section{Introduction}\label{sec:intro}

Understanding the nature of dark matter is one of the most pressing issues in modern astrophysics, cosmology, and particle physics. Indirect detection of Standard Model particles produced in dark matter annihilations or decays is one way to gain an understanding of dark matter (DM) particles and how they may interact with familiar matter. Neutrino telescopes have the ability to detect and/or constrain neutrinos produced in this manner, especially decay of dark matter particles captured within the Sun or Earth \cite{Freese:1985qw,Krauss:1985aaa,Gould:1987ir,Griest:1986yu,Gould:1991hx,Faraggi:1999iu,Arguelles:2019ouk,Cline:2019snp,Albuquerque:2000rk,Albuquerque:2002bj,Bruch:2009rp,Bhattacharya:2019ucd,Mukherjee:2019seu,Anchordoqui:2021dls,Guepin:2021ljb,Xu:2020lrn,Garani:2021feo,Reno:2021cdh}. So far, Super Kamiokande \cite{Super-Kamiokande:2004pou}, AMANDA \cite{AMANDA:2006jtl} and IceCube \cite{IceCube:2016aga} have published limits on localized decay for dark matter particle masses around $\sim 1 - 10^4$ GeV.

Some work has considered decays of more massive particles \cite{Murase:2012xs,Esmaili:2013gha,Murase:2015gea}, including as a possible explanation for the anomalous events seen by ANITA \cite{Anchordoqui:2018ucj,Heurtier:2019git}, and IceCube has also examined limits above $10^4$ GeV \cite{IceCube:2018tkk}. However, recent work suggests that lower-energy secondary neutrinos would have been seen at IceCube, ruling out this explanation \cite{Cline:2019snp,Safa:2019ege}. Models of heavy dark matter include decaying gravitinos \cite{Covi:2008jy} or right-handed neutrinos \cite{Boyle:2018tzc}, while many of the above studies take a phenomenological approach that is agnostic to the particular model involved.

In this paper, we examine the current constraints and reach of IceCube in detecting high-energy neutrino decays from heavy dark matter particles trapped within the Earth, connecting IceCube event rates to dark matter properties, without restricting ourselves to the part of parameter space relevant to the ANITA events. We include a full treatment of tau neutrino propagation and regeneration \cite{Halzen:1998be,Becattini:2000fj,Beacom:2001xn,GonzalezGarcia:2005xw,Reno:2019jtr,Kochocki:2020iie} within the Earth, using a modification of NuTauSim \cite{Alvarez-Muniz:2017mpk}, and give a semi-analytic expression for the resulting spectra.

In particular, we examine the phenomenon of large tau lepton energy loss during propagation and emphasize the role it plays in setting constraints. Our work extends the pre-IceCube results of \cite{Albuquerque:2000rk}, and we use numerical and semi-analytic calculations to provide a simple parametrization of the energy spectrum that will allow straightforward estimation of the effect of future observational developments on these constraints. Our work complements another recent paper \cite{Reno:2021cdh}; in contrast with that work, we use a DM distribution localized very close to Earth's center based on the recent modeling of \cite{Acevedo:2020gro}, we examine both decays and annihilations, and present constraints in terms of dark matter-nucleon cross sections and thermally-averaged annihilation cross sections, to allow for comparison with specific dark matter models.

The structure of the paper is as follows. \sref{sec:model} describes the models and parameters that we investigate, and introduces the formalism that we use to describe the evolution of neutrino flux. \sref{sec:propagation} describes the simulation and presents our numerical and semi-analytic results for high-energy neutrino propagation originating within the Earth. \sref{sec:constraints} uses the results of \sref{sec:propagation} to examine the neutrino spectrum that would be seen at IceCube, as a function of the dark matter parameters. We use this to draw current constraints on the parameter space in comparison with direct detection bounds. Finally, in \sref{sec:conclusion} we summarize our results, discuss implications and dependence on underlying assumptions, and consider possible extensions to future work.

\section{Models and Parameter Space}\label{sec:model}

For greatest generality of our results, we consider two general scenarios that are commonly studied \cite{Arguelles:2019ouk}, and make minimal a priori assumptions about values of parameters or model-specific relationships between them. The general scenarios are
\begin{enumerate}
	\item Decay: Examples include a heavy right-handed neutrino, $\nu_{\rm R}$, which can decay into a tau neutrino plus a Higgs \cite{Boyle:2018tzc,Anchordoqui:2018ucj}, or a decaying gravitino \cite{Covi:2008jy}. We follow most studies in remaining agnostic to the particular model, considering the dark matter particle to have decay time $\tau_{\chi}$.
	\item Annihilation: Dark matter particles annihilate via $X + X \rightarrow \nu + \nu$ (see review \cite{Arguelles:2019ouk}), with thermally-averaged cross section $\langle \sigma v \rangle$ a constant in $v$, i.e. s-wave annihilation.
\end{enumerate}
We assume that there is either a direct decay to neutrinos (as in the above examples) or that any decay chain occurs quickly enough that we can reasonably approximate the neutrino source as the decay location. Our numerical results in \sref{sec:propagation} suggest that variations of $\lesssim 100$ km do not significantly affect our conclusions. A longer-lived mediator would affect our results by reducing the number of neutrinos detected for given dark matter parameters, a point we return to in \sref{sec:conclusion}.

Photons or charged leptons would generically accompany neutrino emission, but we do not consider potential observables related to these. We consider a spin-independent DM-nucleon cross section $\sigma_{\chi N}$ which scales as $A^4$, and in the case of decay, we take $m_{\chi}, \tau_{\chi}, \sigma_{\chi N}$ as independent free parameters, while for annihilation these are $m_{\chi}, \langle \sigma v \rangle, \sigma_{\chi N}$. We do not restrict ourselves to the assumption that dark matter was produced thermally in the early universe, as non-thermal relics evade the unitarity bound that would otherwise preclude $m_{\chi} \gtrsim 10^5$ GeV \cite{Griest:1989wd}.

We now describe the process of dark matter capture within Earth. First, we estimate annihilation and decay rates from the number density and distribution of dark matter particles within Earth. As Earth's galactic orbit passes through dark matter of density $\rho \simeq 0.3$ GeV/cm$^3$ at $v \approx 220$ km/s, particles fall into Earth's potential well and lose energy via collisions with nuclei, becoming trapped with $v < v_{\rm esc}$ (see \cite{Acevedo:2020gro} for a recent calculation). The captured particles thermalize on a timescale much shorter than the age of Earth, and settle into an equilibrium distribution which we represent in terms of the number density as
\begin{align}
	n(t,\vec{x}) & = n_0(t) \exp(-r / r_0),
\end{align}
for constant $r_0$ and central number density $n_0(t)$. At any given time $n_0$ is normalized to the total number $N(t)$ of trapped particles. In particular, $N = \int d^3x \, n(t,\vec{x})$ gives $n_0 = N(t) / (8\pi r_0^3)$. Using $d\Gamma / dV = \langle \sigma v \rangle n(t,\vec{x})^2$, the annihilation rate is then \cite{Griest:1986yu,Albuquerque:2000rk,Bruch:2009rp}
\begin{align}
	\Gamma & = \int d^3x \langle \sigma v \rangle n(t,\vec{x})^2 \, = \, \langle \sigma v \rangle N(t)^2 / (64 \pi r_0^3).
\end{align}
The number of captured particles may be approximated as \cite{Griest:1986yu}
\begin{align}
	N(t) & = N_{\rm eq} \tanh\left( t / \tau \right),
\end{align}
where the eventual equilibrium number is $N_{\rm eq} = (C / C_A)^{1/2}$ and the timescale $\tau = (C  C_A)^{-1/2}$, $C$ is the capture rate, and $C_A \equiv \Gamma / N^2 = \langle \sigma v \rangle / (64\pi r_0^3)$ in our case. The age of Earth is significantly less than $\tau$ \cite{Bruch:2009rp}, so we can expand $\tanh(t/\tau) \approx (t/\tau)$, leading to $N(t) \simeq Ct$, so the annihilation rate is
\begin{align}\label{eq:ann-rate-v1}
	\Gamma_{\rm ann.} & \simeq \frac{\langle \sigma v \rangle C^2 t^2}{64 \pi r_0^3}.
\end{align}
The extent $r_0$ of the thermalized dark matter distribution may be estimated as \cite{Acevedo:2020gro}
\begin{align}\label{eq:thermalization-radius}
	r_0 & \simeq \left( 2 \, {\rm km}\right) \left( \frac{10^7 \, {\rm GeV}}{m_{\chi}} \right)^{1/2}
\end{align}
when at Earth's core the temperature is 5000 K and density is 10 g/cm$^3$. We adopt these nominal values and therefore do not include slowly-varying corrections that would account for different values. Furthermore, Ref. \cite{Acevedo:2020gro} estimates the capture rate for isotope-dependent $(A^4)$ DM-nucleon scattering as
\begin{align}\label{eq:capture-rate}
	C_{\chi N} & = \begin{cases} (2.45 \times 10^{22} \, {\rm s}^{-1}) \left( \frac{10^3 \, {\rm GeV}}{m_{\chi}} \right) \, & \, {\rm for} \, \left( \frac{m_{\chi}}{1.66\times 10^{12} \, {\rm GeV}} \right) < \left( \frac{\sigma_{\chi N}}{10^{-26} \, {\rm cm}^2} \right) \\ (8.74 \times 10^{27} \, {\rm s}^{-1}) \left( \frac{10^8 \, {\rm GeV}}{m_{\chi}} \right)^{7/2} \left( \frac{\sigma_{\chi N}}{10^{-26} \, {\rm cm}^2} \right)^{5/2} & \, {\rm for } \, \left( \frac{m_{\chi}}{1.66\times 10^{12} \, {\rm GeV}} \right) > \left( \frac{\sigma_{\chi N}}{10^{-26} \, {\rm cm}^2} \right) \end{cases},
\end{align}
where capture is efficient for large enough cross section but drops rapidly below a certain value that depends on the mass. So, finally, we can combine \eref{eq:ann-rate-v1} with \eref{eq:thermalization-radius} and \eref{eq:capture-rate}, including a factor of 2 to account for each annihilation producing two neutrinos, to find:
\begin{align}\label{eq:ann-rate-v2}
	\Gamma_{\rm ann.} & \simeq \begin{cases} (1.50 \times 10^{28} \, {\rm s}^{-1}) \left( \frac{10^7 \, {\rm GeV}}{m_{\chi}} \right)^{1/2} \left( \frac{\langle \sigma v \rangle}{10^{-25} \, {\rm cm}^3{\rm s}^{-1}} \right) \, & \, {\rm for} \, \left( \frac{m_{\chi}}{1.66\times 10^{12} \, {\rm GeV}} \right) < \left( \frac{\sigma_{\chi N}}{10^{-26} \, {\rm cm}^2} \right) \\ (1.91 \times 10^{54} \, {\rm s}^{-1}) \left( \frac{\langle \sigma v \rangle}{10^{-25} \, {\rm cm}^3{\rm s}^{-1}} \right) \left( \frac{10^7 \, {\rm GeV}}{m_{\chi}} \right)^{11/2} \left( \frac{\sigma_{\chi N}}{10^{-26} \, {\rm cm}^2} \right)^5 &  \, {\rm for} \, \left( \frac{m_{\chi}}{1.66\times 10^{12} \, {\rm GeV}} \right) > \left( \frac{\sigma_{\chi N}}{10^{-26} \, {\rm cm}^2} \right) \end{cases},
\end{align}
where we have used $t = 4.5$ Gyr as the age of Earth. Our constraints in subsequent sections will take the DM-nucleon cross section $\sigma_{\chi N}$ and the thermally-averaged DM self-annihilation cross section $\langle \sigma v \rangle$ to vary as independent parameters.

For the case of dark matter particles that decay with lifetime $\tau_{\chi}$, the decay rate is
\begin{align}
	\Gamma_{\rm decay} & = \frac{1}{\tau_{\chi}}N(t) \, \simeq \, \frac{1}{\tau_{\chi}} C t.
\end{align}
Again using capture rate estimate \eref{eq:capture-rate}, we find
\begin{align}\label{eq:decay-rate}
	\Gamma_{\rm decay} & \simeq \begin{cases} (2.45 \times 10^{22} \, {\rm s}^{-1}) \left( \frac{10^3 \, {\rm GeV}}{m_{\chi}} \right) \left( \frac{t_{\rm Earth}}{\tau_{\chi}} \right) \, & \, {\rm for} \, \left( \frac{m_{\chi}}{1.66\times 10^{12} \, {\rm GeV}} \right) < \left( \frac{\sigma_{\chi N}}{10^{-26} \, {\rm cm}^2} \right) \\ (8.74 \times 10^{27} \, {\rm s}^{-1}) \left( \frac{10^8 \, {\rm GeV}}{m_{\chi}} \right)^{7/2} \left( \frac{\sigma_{\chi N}}{10^{-26} \, {\rm cm}^2} \right)^{5/2} \left( \frac{t_{\rm Earth}}{\tau_{\chi}} \right) & \, {\rm for } \, \left( \frac{m_{\chi}}{1.66\times 10^{12} \, {\rm GeV}} \right) > \left( \frac{\sigma_{\chi N}}{10^{-26} \, {\rm cm}^2} \right) \end{cases}.
\end{align}
We will take $\sigma_{\chi N}$ and $\tau_{\chi}$ to be independent parameters in the analysis that follows.

Formally, we can express the neutrino flux arriving at a detector in terms of the observed energy $E$ and the initial energy $\tilde{E}$ as
\begin{align}
	\frac{d\Phi}{d E} & = 2\pi \int r^2 d\cos\theta \int d\tilde{E} \frac{d \Gamma}{dV}(r,\theta) f(E, \tilde{E}, x) h(\tilde{E})
\end{align}
where $d\Gamma / dV$ is the local decay or annihilation rate (defined by requiring $\Gamma = \int dV (d\Gamma / dV)$ with \eref{eq:ann-rate-v2} or \eref{eq:decay-rate}), $r$ and $\theta$ are radial and polar coordinates in detector-centered spherical coordinates, and we have assumed azimuthal symmetry. The function $h(\tilde{E})$ gives the initial spectrum of decays, and $f(E,\tilde{E},x)$ is the probability for a neutrino of initial energy $\tilde{E}$ to arrive at the detector with energy $E$, traversing column depth $x \equiv \int dz \rho(z)$, where $z$ is position along line of sight and $\rho(z)$ is local matter density. Due to the expected localization of heavy dark matter particles around Earth's center as in \eref{eq:thermalization-radius}, we approximate all decays as occurring at the center, simplifying the above expression. The validity of this is further examined in \sref{sec:propagation}. We will see that the initial energy $\tilde{E}$ is irrelevant above $\sim 10^7$ GeV, and therefore the form of $h(\tilde{E})$ will be irrelevant to most of our parameter space. In \sref{sec:propagation}, we numerically study evolution of neutrino flux through Earth in order to determine $f(E,\tilde{E})$.

\section{Propagation of Tau Neutrinos Within the Earth}\label{sec:propagation}

As tau neutrinos propagate through the Earth, they undergo neutral current (NC) and charged current (CC) interactions with the surrounding rock. CC interactions occur approximately 72\% of the time, and result in a shower of particles, including a tau lepton, which keeps some of the neutrino’s energy. As it moves, the tau loses energy due to the photonuclear, Bremsstrahlung, and pair production effects (the first of which dominates at energies above $\sim 10^{8}$ GeV), as a function of its current energy and the density of its surroundings. The particle eventually decays back into a tau neutrino, a stochastic event which occurs after the particle travels a randomized multiple of the tau decay length; this is tau neutrino regeneration \cite{Halzen:1998be,Becattini:2000fj,Beacom:2001xn,GonzalezGarcia:2005xw,Reno:2019jtr,Kochocki:2020iie}. We approximate the energy loss resulting from both the production and the decay of the $\tau$ as 30\%. Before analyzing our numerical results, we discuss the energy loss of the tau leptons analytically to better illustrate certain features of the full results. 

\subsection{Semi-Analytic Estimates}\label{sec:semi-analytic}

\begin{figure}[t]
	\centering
	\begin{subfigure}[b]{0.445\textwidth}
		\centering
		\includegraphics[width=\textwidth]{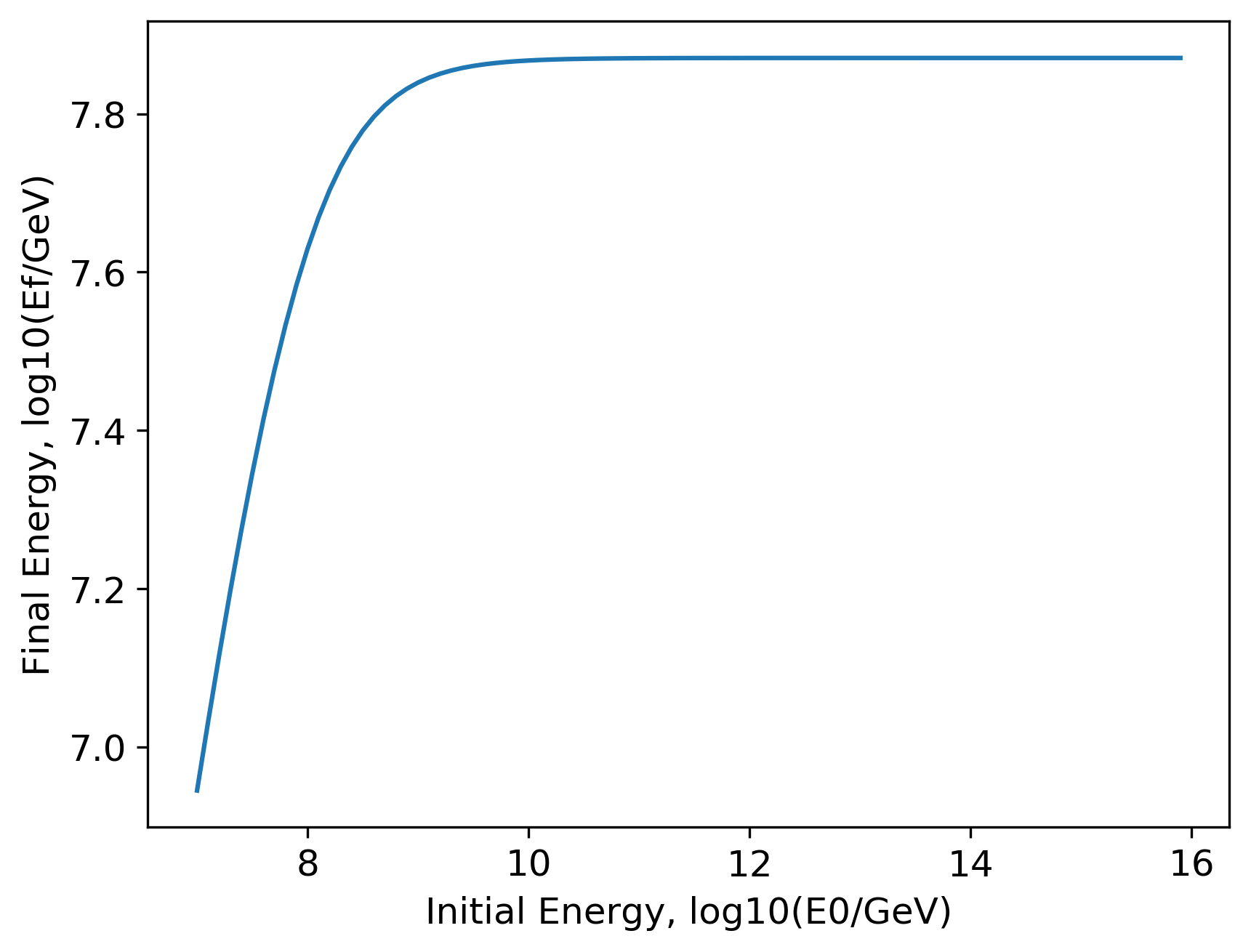}
		\caption{\label{fig:analytic-eloss}}
	\end{subfigure} \, \, \, \, \, \, %
	\begin{subfigure}[b]{0.455\textwidth}
		\centering
		\includegraphics[width=\textwidth]{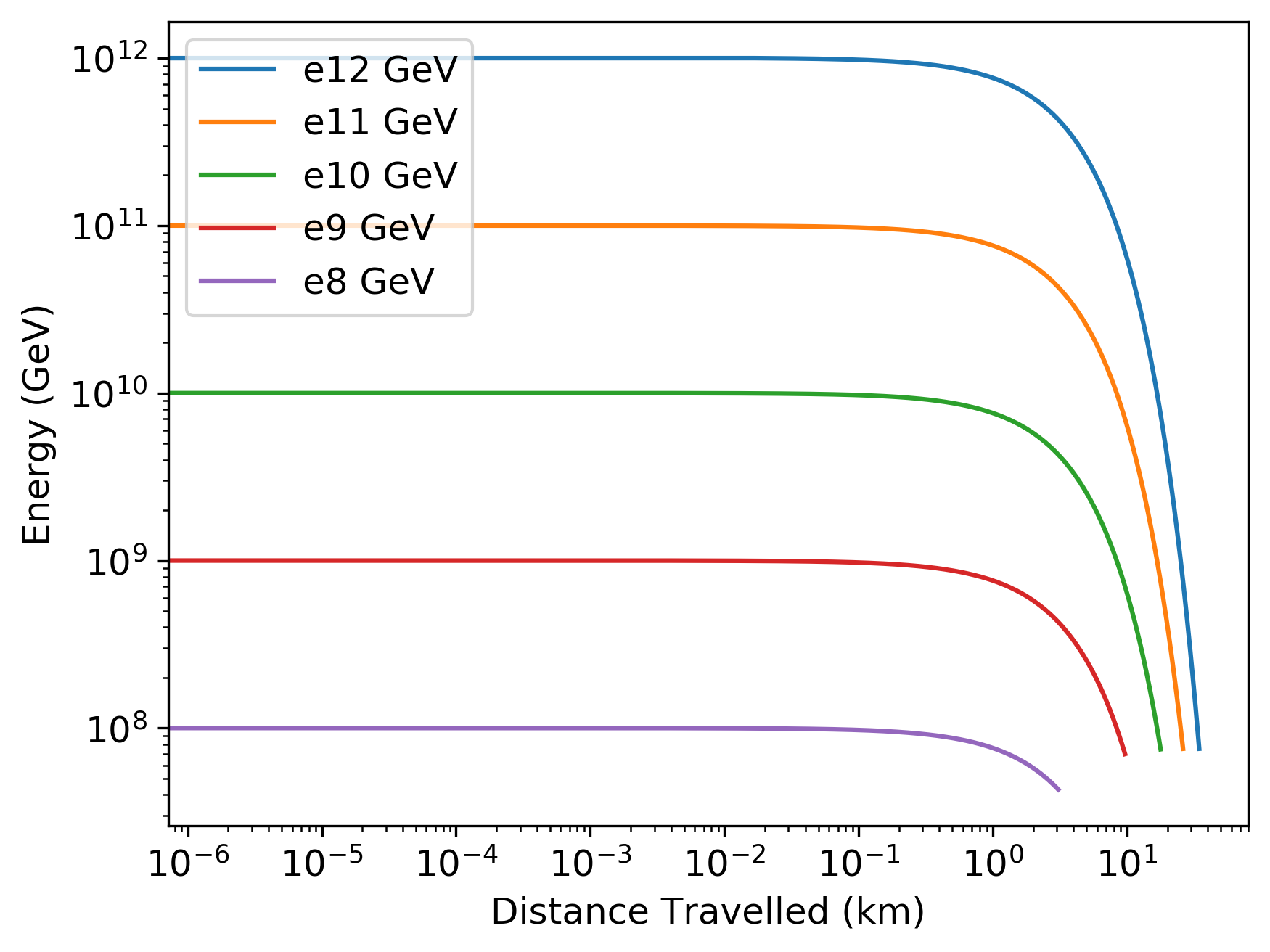}
		\caption{\label{fig:tau-eloss}}		
	\end{subfigure}
\caption{\label{fig:tau-results}Fig. \ref{fig:analytic-eloss} : The final energy of a tau lepton at the moment of decay as a function of its initial energy, using estimates \eref{eq:eloss-integrated} and \eref{eq:tau-dist-solved}.  Fig.~\ref{fig:tau-eloss}: Tau lepton energies as they move through rock, for starting values between $10^{8} $ and $10^{12}$ GeV. They travel exactly one lifetime, all decaying with energy $\sim 10^8$ GeV having travelled $\mathcal{O}(10^1 \, {\rm km})$.}
\end{figure}

Since the photonuclear effect accounts for most energy loss above $\sim 10^{8}$ GeV, we can describe the energy loss rate of a tau lepton of energy $E$ as
\begin{align}\label{eq:eloss-original}
	\left \langle \frac{dE} {dx} \right \rangle & \simeq - \beta(E) E,	
\end{align}
where $\beta$ is parametrized by
\begin{align}\label{eq:beta-parametrization}
	\beta (E) = p_0 + p_1(E/GeV)^{p_2},
\end{align}
with $p_0 = 2.06 \times 10^{-7}$ cm$^2$/g, $p_1 = 4.93 \times 10^{-9}$ cm$^2$/g, $p_2 = 0.228$ in the ALLM parametrization \cite{Abramowicz:1997ms,Alvarez-Muniz:2017mpk}. The exponent causes $\beta$ to vary slowly with energy, and in this section we approximate it by its value at $E = 10^8$ GeV, i.e. we take $\beta \simeq 5.0 \times 10^{-7}$ cm$^2$ / g. Note that our numerical results use the full energy dependence. Integrating \eref{eq:eloss-original} under this approximation, we find
\begin{align}\label{eq:eloss-integrated}
	E = E_{i} e^{-\beta \rho z},
\end{align}
where $\rho$ is the average density along the tau's path, $z$ is the distance travelled by the particle, and $E_i$ is the initial energy of the tau lepton. Since the total energy loss depends on the distance travelled, which depends on the particle's lifetime and velocity, the distance is also dependent on the particle's energy. As an approximation, we assume the tau decays after exactly one lifetime ($\tau$, or $\tau_0$ in its rest frame), and express this as
\begin{align}\label{eq:tau-decay-1}
	1 & = \int_0^{t_{\rm dec.}} \frac{1}{\tau} dt \, = \, \int_0^{z_{\rm dec.}} \frac{mc^2}{E} \frac{1}{c\tau_0} dz,
\end{align}
where $m$ is the tau mass and $z_{\rm dec.}$ represents the distance travelled during the tau's lifetime. Using \eref{eq:eloss-integrated} for the energy, evaluating the $z$ integral and solving for the distance gives
\begin{align}\label{eq:tau-dist-solved}
	z_{\rm dec.} = \frac{1}{\beta \rho} \ln{\left(\frac{\beta \rho ct_{0}E_0}{mc^{2}} + 1 \right)}.
\end{align}
This result can be used with \eref{eq:eloss-integrated} to show the tau's energy just before it decays, across a range of starting energies (as seen in \ref{fig:analytic-eloss}); this shows that tau leptons above $\sim 10^{8}$ GeV lose energy until reaching approximately $10^{7.8}$ GeV, regardless of their starting energy. Above $\sim 10^{8}$ GeV the effect is dramatic, but approaching and below that point the energy losses become negligible. This can also be seen in \ref{fig:tau-eloss}: no matter their starting energy, the particles slow down dramatically within a short distance (relative to the path length of the tau lepton and neutrino) until reaching $\sim 10^{7.8}$ GeV, after which point they lose insignificant amounts of energy until decaying.

The stochastic nature of real decays means that the full numerical simulation in \sref{sec:numerical-results} will find a distribution extending above and below a central value. Furthermore, the estimate of this section only considers propagation of a tau during one lifetime; the results of \sref{sec:numerical-results} also account for energy losses during decays and more than one instance of tau propagation along the path, so they result in typical energies at the surface being lower. However, the phenomenon shown here - where high-energy tau propagation drives widely-varying initial energies toward similar final energies - does carry over.

\subsection{Numerical Results}\label{sec:numerical-results}

\begin{figure}[t]
	\centering
	\begin{subfigure}[b]{0.45\textwidth}
		\centering
		\includegraphics[width=\textwidth]{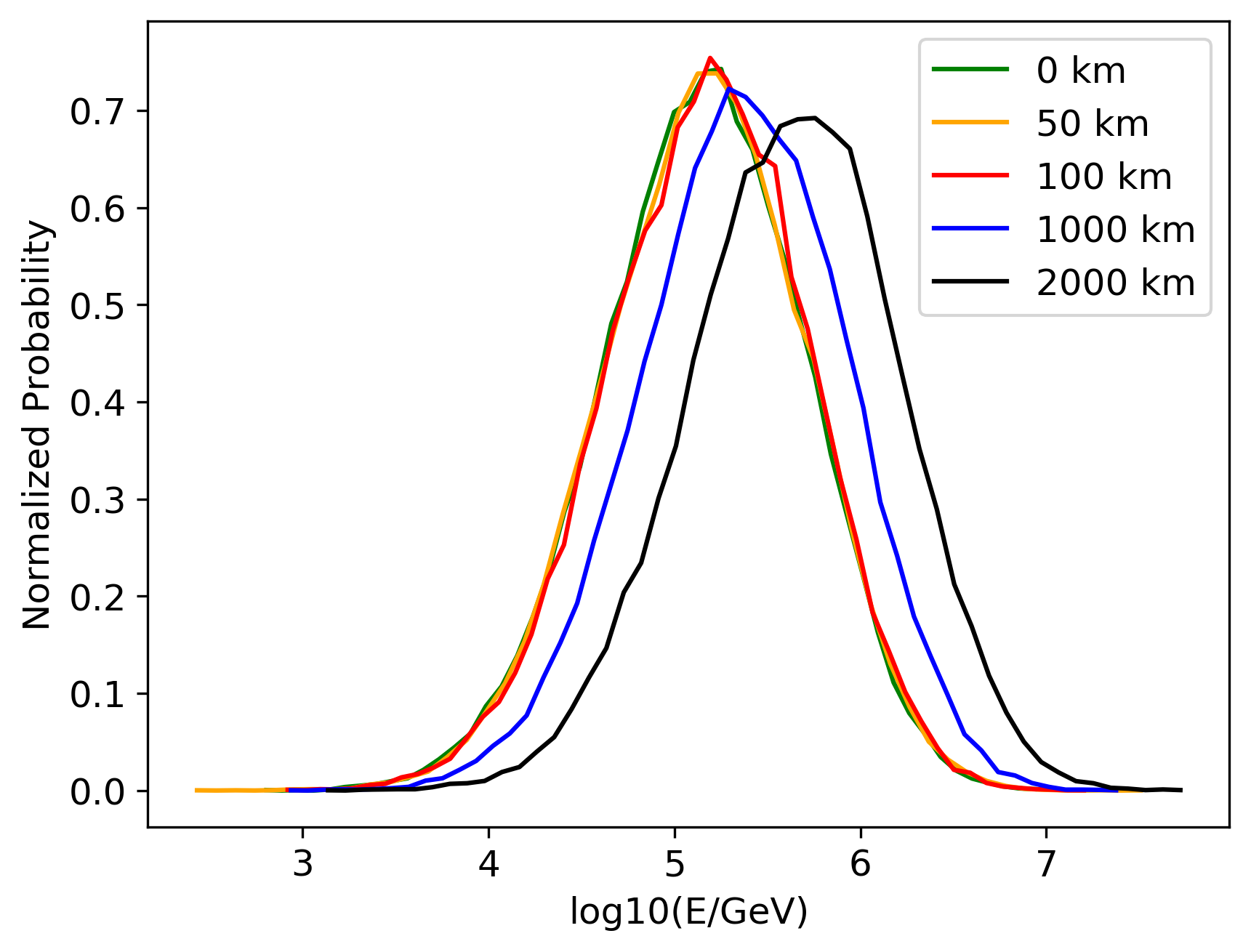}
		\caption{\label{fig:starting-offset}}
	\end{subfigure} \, \, \, \, \, \, %
	\begin{subfigure}[b]{0.45\textwidth}
		\centering
		\includegraphics[width=\textwidth]{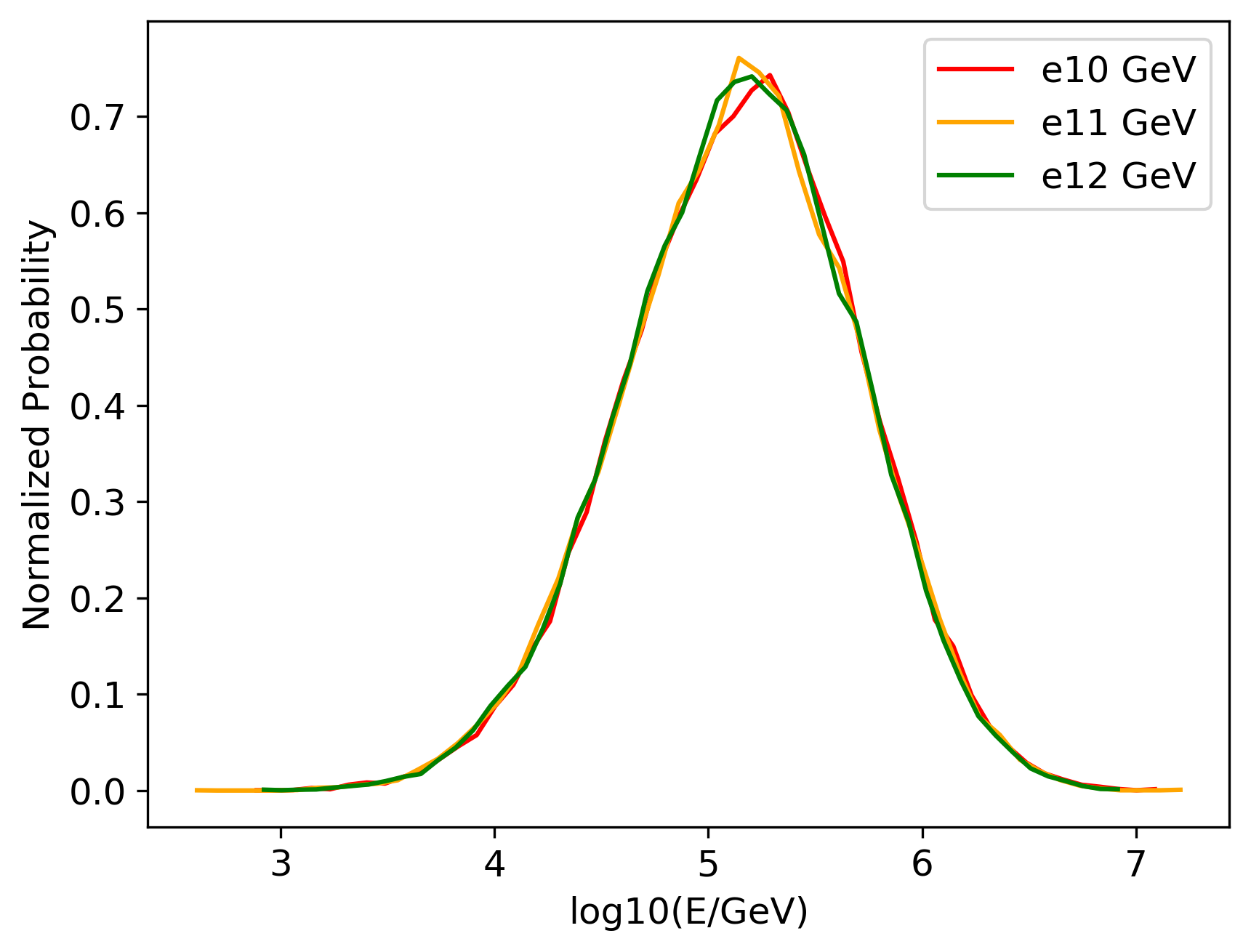}
		\caption{\label{fig:histogram}}
	\end{subfigure}
\caption{\label{fig:final-energy} Fig.~\ref{fig:starting-offset} shows the final energy distribution of $10^{12}$ GeV neutrinos generated some distance from the center, along the line of sight. \\ Fig.~\ref{fig:histogram} shows the final distribution of tau neutrinos generated at three different energies. Due to the effects in \fref{fig:tau-eloss}, the means and standard deviations of every distribution are within 5\% of each other for energies between $10^{7}$ and $10^{17}$ GeV.  }
\end{figure}

To produce an expected distribution of tau neutrino energies at the surface of the Earth, we use a modification of NuTauSim \cite{Alvarez-Muniz:2017mpk}. The Earth's density profile is modeled using the Preliminary Earth Reference Model \cite{Dziewonski:1989}. Since the dark matter is localized around the center of the Earth (\eref{eq:thermalization-radius}) we approximate it to the exact center. That this is accurate can be seen in \fref{fig:starting-offset}. Most of the dark matter is located well within 100 km of the center, and a 100 km displacement has a negligible effect on the propagation. Only at extreme distances, where effectively no neutrinos are produced, does the displacement matter, so this is a good approximation.

The energy losses result in final energies at the Earth's surface being, on average, around $10^{5}$ GeV, for all starting energies above $\sim 10^{8}$ GeV. This can be seen in \ref{fig:histogram}, where each $f(E,\tilde{E})$ is given as a normalized probability density function (PDF) representing an average over $\sim 50,000$ neutrinos numerically propagated from the center of the Earth. Since these distributions are nearly identical (for energies as high as $10^{17}$ GeV), they can be described with the following normalized probability distribution in terms of $\epsilon \equiv \log_{10}({\rm E / GeV})$:
\begin{align}\label{eq:gaussian}
	f(\epsilon) = \frac{1}{\sigma \sqrt{2\pi}} e^{-\frac{1}{2}\left (\frac{\mu-\epsilon}{\sigma}\right )^2},
\end{align} 
where $\sigma = 0.541$ and $\mu = 5.148$.\footnote{So $f(E) = \left[ \sqrt{2\pi} \sigma \ln(10) E \right]^{-1} \exp\left[ (\log_{10}({\rm E / GeV}) - \mu)^2 / (2 \sigma^2) \right]$ is the normalized PDF in terms of $E$; this will be convenient in \sref{sec:constraints}.} 

\begin{figure}[t]
	\centering
	\begin{subfigure}[b]{0.45\textwidth}
		\centering
		\includegraphics[width=\textwidth]{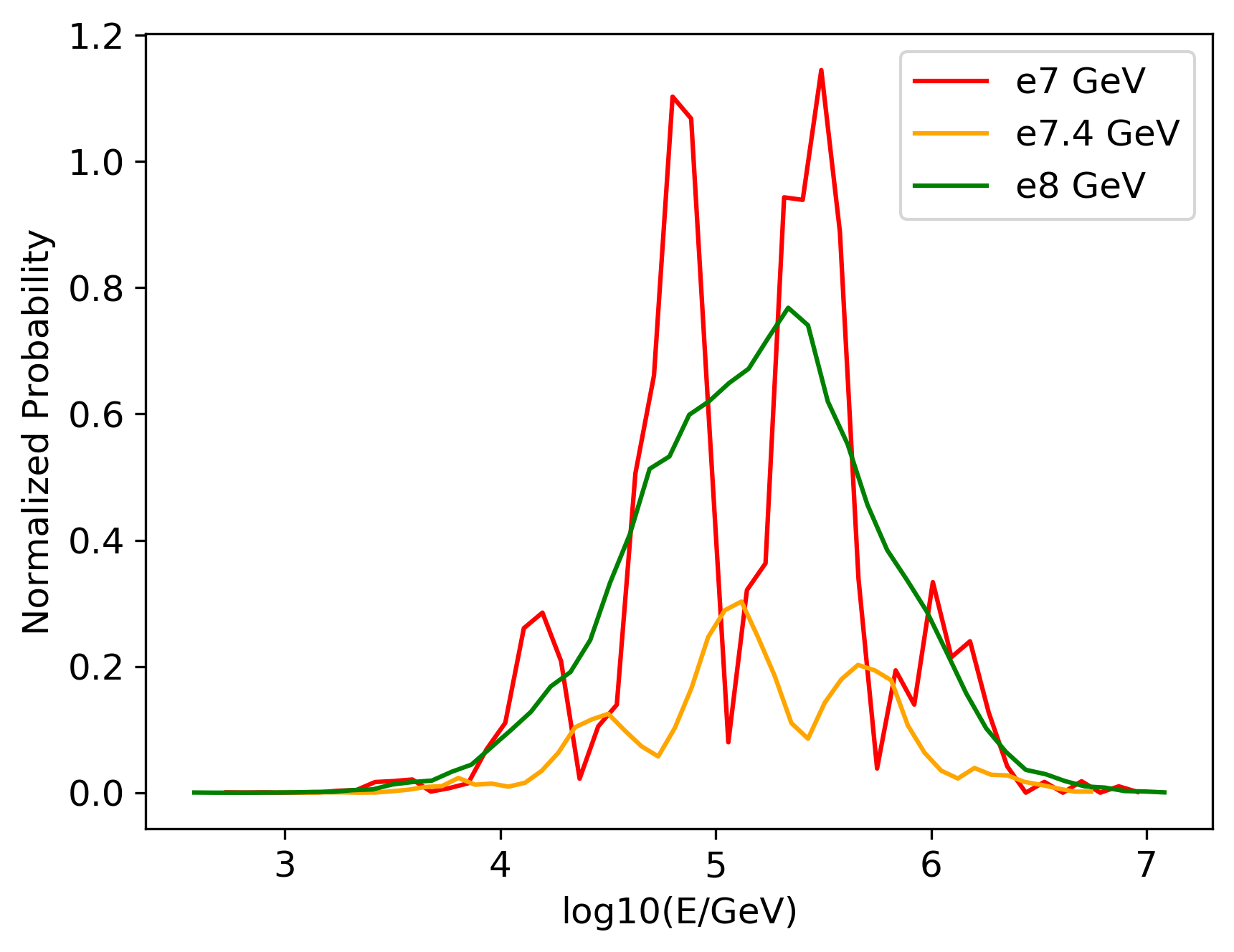}
		\caption{\label{fig:spikes-transition}}
	\end{subfigure} \, \, \, \, \, \, %
	\begin{subfigure}[b]{0.44\textwidth}
		\centering
		\includegraphics[width=\textwidth]{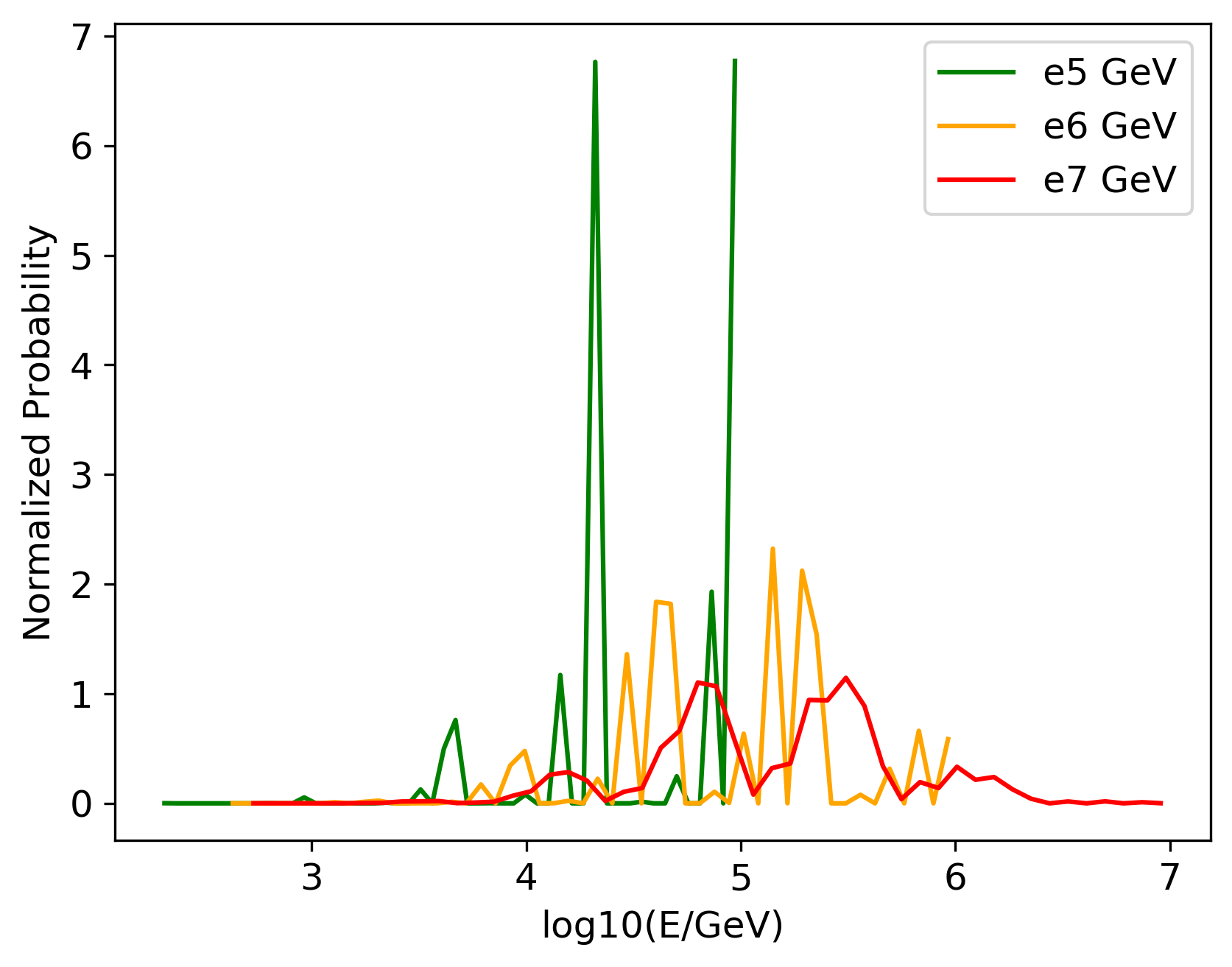}
		\caption{\label{fig:spikes-low-energy}}
	\end{subfigure}
\caption{\label{fig:low-energy-curves} Normalized final energy distributions for neutrinos with varying initial energies.} 
\end{figure}

Below initial energies of $10^{8}$ GeV, this becomes less accurate. The energy losses from the tau propagation stages become negligible next to those of NC and CC interactions, as well as the decay process of the tau leptons into tau neutrinos. As a result, neutrinos tend towards specific energies that depend on the number of interactions they experience, which can be seen in \fref{fig:spikes-transition}. The gaussian gives way to several specific energy levels, an effect that becomes even more pronounced in \fref{fig:spikes-low-energy}. At $10^{5}$ and $10^{6}$ GeV, most neutrinos reach the surface with one of two energies: they either never interact and keep their initial energy, or undergo exactly one CC interaction and reach the surface with about 10\% of their initial energy.

\section{Parameter Reach of IceCube HESE Events}\label{sec:constraints}

As mentioned previously, we expect the general phenomenon described in \sref{sec:propagation} to be important for any observation of ultra-high energy tau neutrinos. We now illustrate this effect by taking as an example the existing IceCube data set for high-energy starting events (HESE) at energies above 60 TeV over 2635 days of live time \cite{IceCube:2020wum,IceCube:2020abv}. To date, the HESE sample consists of 60 events including two double cascade events that have been identified as tau neutrinos. This sample is consistent with a diffuse background with power-law energy dependence. For our analysis we exclude any region of parameter space that predicts 10 IceCube HESE events over 2635 days for energies greater than 60 TeV. We chose this as an estimate of a number of events that would be clearly localized compared with the 60 total events, and any $\mathcal{O}(10)$ number leads to essentially the same results. We also estimate the IceCube-Gen2 sensitivity following \cite{IceCube-Gen2:2020qha}.

\begin{figure}[t]
	\centering
	\begin{subfigure}[b]{0.49\textwidth}
		\centering
		\includegraphics[width=\textwidth]{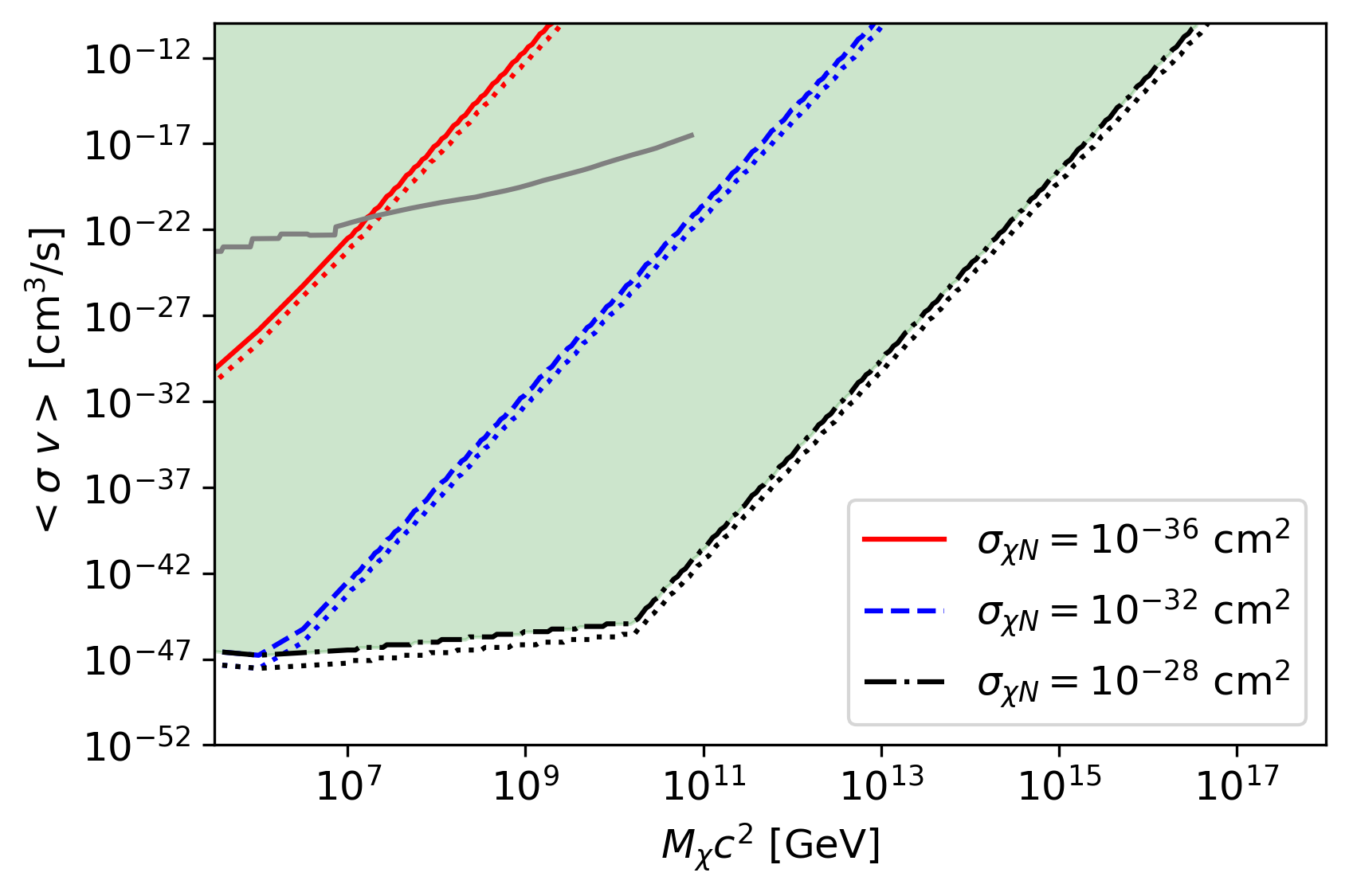}
		\caption{\label{fig:ann-si}}
	\end{subfigure} \, \, %
	\begin{subfigure}[b]{0.49\textwidth}
		\centering
		\includegraphics[width=\textwidth]{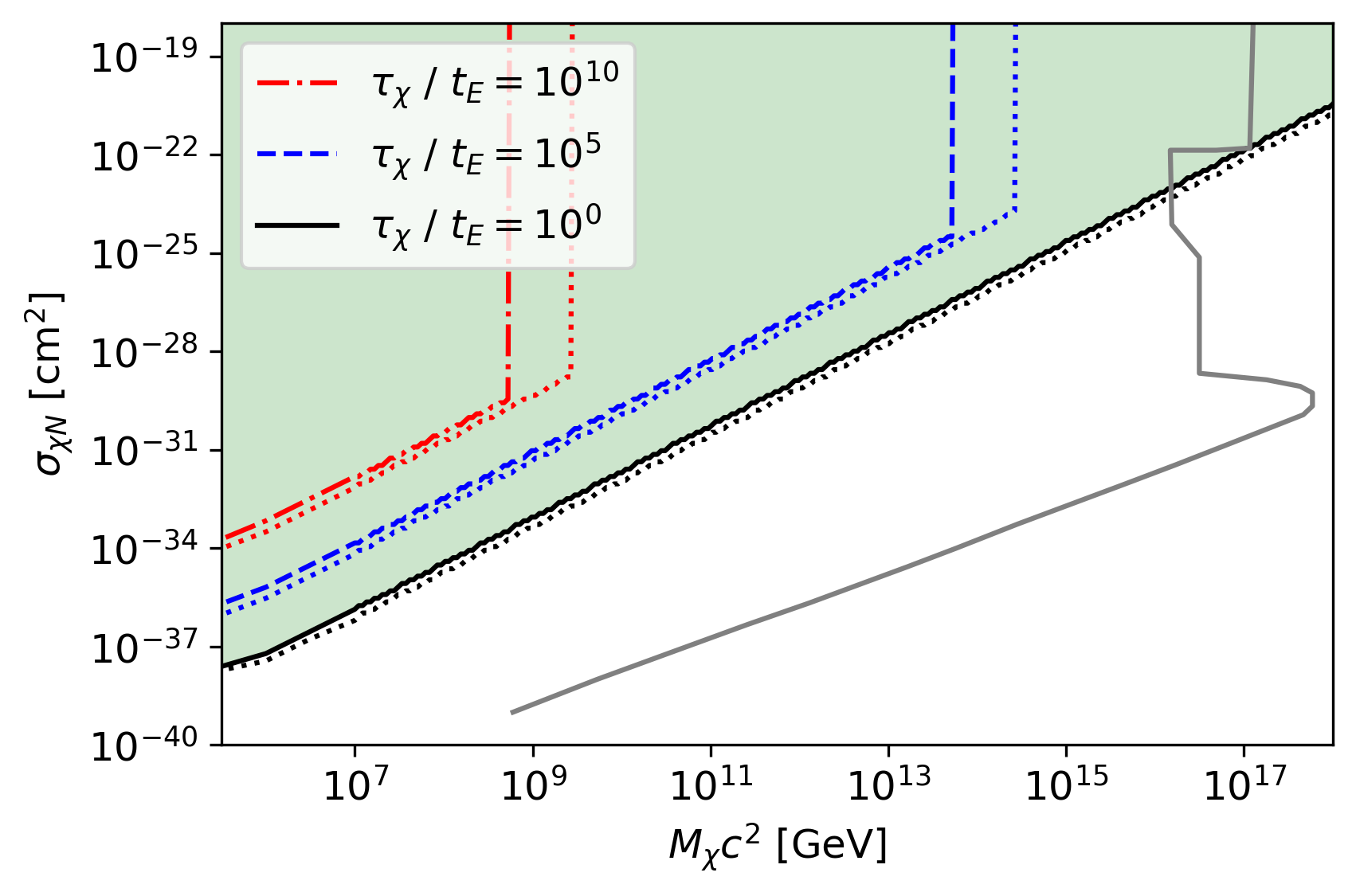}
		\caption{\label{fig:decay-si}}
	\end{subfigure}
\caption{\label{fig:constraints} Parameter space reach of IceCube HESE events (dotted lines: IceCube-Gen2) for dark matter self-annihilation (Fig.~\ref{fig:ann-si}) and decay (\fref{fig:decay-si}), for spin-independent ($A^4$-scaling) DM-nucleon cross sections. The gray line in \fref{fig:ann-si} represents existing constraints on $\langle \sigma v \rangle$ \cite{Arguelles:2019ouk}; the gray line in \fref{fig:decay-si} shows $\sigma_{\chi N}$ constraints from \cite{Clark:2020mna}. In \fref{fig:decay-si} the DM decay time is given as $\tau_{\chi} / t_{\rm E}$, where $t_{\rm E} = 4.5 \times 10^9$ years is the age of Earth. In \fref{fig:ann-si} we see that the $\sigma_{\chi N}$ values needed to evade the bound we have computed for a given annihilation cross section are in tension with existing constraints on $\sigma_{\chi N}$ (compare with solid gray line in \fref{fig:decay-si}).} 
\end{figure}

Formally, the number of events observed at IceCube during time $T$ is
\begin{align}\label{eq:n-events}
	N & = T \int dE \int d\tilde{E} \, f(E,\tilde{E}) \, \frac{d\Gamma}{d\tilde{E}}\frac{1}{4\pi R_{\oplus}^2} \, A_{\rm eff}(E),
\end{align}
where (as before) $E$ is the neutrino energy at the detector and $\tilde{E}$ is the energy at production (time of decay), and $A_{\rm eff}$ is the IceCube detector effective area for tau neutrinos (see e.g. \cite{IceCube:2013low}). For simplicity we approximate the decay spectrum of the dark matter particle with a delta function at 0.1$m_{\chi}c^2$, although recalling the discussion of \sref{sec:semi-analytic}, the results will be almost independent of assumptions about this spectrum. We describe propagation using the results of \sref{sec:propagation} as encoded in \eref{eq:gaussian}, which leaves us with
\begin{align}\label{eq:n-events-v2}
	N & = T \frac{\Gamma}{4\pi R_{\oplus}^2} \int_{60 \, {\rm TeV}}^{\infty} dE \, f(E,\tilde{E}) \, A_{\rm eff}(E).
\end{align}
We use \eref{eq:ann-rate-v2} or \eref{eq:decay-rate} for annihilation or decay rates, use the IceCube effective areas for tau neutrino detection \cite{IceCube:2013low,IceCube:2020wum}, and evaluate the integral \eref{eq:n-events-v2} numerically using the results of \sref{sec:propagation}. Sampling the parameter space of $m_{\chi}$, $\sigma$ we find constrained regions shown in \fref{fig:constraints}. \fref{fig:ann-si} shows the case of dark matter self-annihilation, while \fref{fig:decay-si} shows the case of dark matter decay. In both cases, projected IceCube-Gen2 limits are included as dotted lines. Note that despite the formal upper limit on the integral in \eref{eq:n-events-v2}, the gaussian tail from \eref{eq:gaussian} causes the integrand to rapidly drop by $10^7$ GeV, so the result is fairly insensitive to details of IceCube's highest energy reach.

For decays above $\sim 10^7$ GeV, the variation of the constraint curve with mass is entirely due to the annihilation rate, \eref{eq:ann-rate-v2}, or decay rate, \eref{eq:decay-rate}. This is clear in \eref{eq:n-events-v2}, where the form of $f(E,\tilde{E})$ in \eref{eq:gaussian} causes the integral to have no dependence on the initial energy or mass scale. This effect makes the features of the plot above $\sim 10^7$ GeV easy to understand. First, we use \eref{eq:gaussian} to numerically evaluate
\begin{align}\label{eq:constraint-int}
	\int_{60 \, {\rm TeV}}^{\infty} dE \, f(E) \, A_{\rm eff}(E) & \simeq 5.1 \, {\rm m}^2.
\end{align}
The result of \eref{eq:constraint-int} now allows us to evaluate \eref{eq:n-events-v2} analytically for cases where the initial neutrino energy is above $\sim 10^7$ GeV:
\begin{align}\label{eq:n-events-v3}
	N & \simeq T \, \Gamma \frac{5.1 \, \rm m^2}{4\pi R_{\oplus}^2} \, \simeq \, 10^{-14} \, T \, \Gamma,
\end{align}
obtaining $\Gamma$ from \eref{eq:ann-rate-v2} or \eref{eq:decay-rate} for annihilation or decay, respectively.

To place the annihilation results (\fref{fig:ann-si}) in context, we note that searches for gamma ray and neutrino signals from DM annihilation at the galactic center limit $\langle \sigma v \rangle$ to less than about $10^{-24}$ to $10^{-26}$ cm$^3$ s$^{-1}$ between $10^2$ and $10^5$ GeV \cite{Fermi-LAT:2015qzw,ANTARES:2019svn,Mazziotta:2020zte}, while in our region of interest above $10^5$ GeV, constraints are somewhat weaker at roughly $10^{-25}$ to $10^{-20}$ cm$^3$/s \cite{Arguelles:2019ouk}, which we show in \fref{fig:ann-si}. In \fref{fig:decay-si} we show constraints on the DM-nucleon cross section from \cite{Clark:2020mna}, assuming $A^4$ scaling (see also \cite{Kavanagh:2017cru}). In \fref{fig:decay-si}, the ``elbow" in the constraint is defined by the boundary between the high-cross section and low-cross section scaling behavior in the capture rates \eref{eq:capture-rate}, i.e. it falls along a line defined by $\sigma_{\chi N} / {\rm cm}^2 = 6.02 \times 10^{13} (m_{\chi} / {\rm GeV})$.

There is a clear tension between limits on $\sigma_{\chi N}$ for decays or annihilations of a given mass, and sensitivity to smaller values of $\langle \sigma v \rangle$: the latter requires larger DM-nucleon cross sections $\sigma_{\chi N}$ (so DM particles are trapped in Earth more efficiently), running up against the direct-detection bounds as seen in \fref{fig:decay-si}. As seen in both \fref{fig:ann-si} and \fref{fig:decay-si}, the increased event rate for IceCube-Gen2 should not have a significant qualitative effect on these results.

This parameter tension provides context for the interpretation of other results that constrain decay or annihilation rates. For IceCube to be sensitive to neutrinos from dark matter particles that evade the direct detection bounds, the capture and thermalization process would have to be significantly altered. The effect of any such modifications can be easily checked using our approximation \eref{eq:n-events-v3} for a dark matter distribution localized near Earth's center. We note that these comparisons can depend somewhat on simplifying assumptions, see e.g. \cite{Digman:2019wdm}.

While previous results are not exactly comparable to ours due to different assumptions, we can make some comparisons. For example, decays of dark matter particles with lifetimes on the order of $10^{27}$ to $10^{28}$ s are considered in Ref. \cite{Esmaili:2013gha} (galactic and extragalactic dark matter) and Ref. \cite{Reno:2021cdh} (dark matter decays in Earth). Since $10^{27}$ s $\sim 10^{10} \ t_{\rm E}$, this corresponds approximately to the dot-dashed (red) curve in \fref{fig:decay-si}, where dark matter-nucleon cross sections are well above the direct detection bound. As noted in \sref{sec:intro}, Ref. \cite{Reno:2021cdh} also assumes different dark matter distributions from the thermalized distribution we have taken, considering dark matter to be uniformly distributed within Earth, or to be proportional to the density profile of baryonic matter. Therefore, it is inherently making different assumptions about dark matter properties involved in capture and thermalization.

\section{Summary \& Conclusions}\label{sec:conclusion}

In this paper, we have shown how rates of high-energy tau neutrinos from decaying or annihilating dark matter are related to dark matter properties such as the dark matter-nucleon cross section and decay rate or thermally-averaged annihilation cross section. Given existing constraints on dark matter properties, we find that this scenario does not allow room for high-energy Earth-emerging tau neutrino events at IceCube to be interpreted in terms of decaying or annihilating dark matter. While previous work has cast doubt on such an interpretation of the anomalous ANITA events \cite{Cline:2019snp,Safa:2019ege}, we examine a much broader range of dark matter masses. Our explicit connection between dark matter properties -- constrained experimentally -- and expected IceCube event rates adds perspective to the existing literature on searches for neutrino signatures of dark matter.

Our work included a numerical simulation that showed how significant tau lepton energy losses during tau neutrino regeneration drive a wide range of initial neutrino energies toward a distinct spectrum, as shown in \eref{eq:gaussian} and \fref{fig:histogram}. When combined with estimates of dark matter capture and thermalization in relation to decay or annihilation rates, this allows for a semi-analytic estimate of the IceCube HESE event rate, \eref{eq:n-events-v3}. This result provides a useful way to quickly check the effect of, for example, quite different assumptions about the dark matter capture and thermalization process.

More sensitive dedicated searches for localized ultra-high energy tau neutrinos could be performed directly from future experiments and upgrades \cite{IceCube-Gen2:2020qha,Ishihara:2019aao,POEMMA:2020ykm}. While these could lead to more sensitivity than we have inferred from the public IceCube HESE sample, we don't expect this to change the qualitative picture we have presented. We have also assumed that tau neutrinos are immediate decay products of the dark matter particle. If there is instead a mediator that eventually decays to neutrinos, with a lifetime allowing many mediator particles to escape Earth before decaying, IceCube could still constrain this but with reduced sensitivity that would depend on model assumptions. There are a number of other ways in which Standard Model uncertainties, dark matter model dependence, and detector subtleties enter the above calculation. While these do not change the general features of our conclusions, future analyses or unexpected results in high-energy neutrino physics (e.g. the tau neutrino cross section \cite{Denton:2020jft}) could be cause to revisit this.

\acknowledgments

MS was supported by a summer research fellowship at Bowdoin College. This research made use of the Bowdoin Computing Grid, and we thank Dj Merrill for assistance. We also thank the referees for valuable comments.

\end{document}